%% file: main.tex
\documentclass{article} 
\usepackage{iclr2024_conference,times}

\input{math_commands.tex}

\usepackage{hyperref}
\usepackage{url}
\usepackage{listings}
\usepackage{verbatim}
\usepackage{multirow}
\usepackage{graphicx}
\usepackage{float}

\title{\dataset: Can Large Language Models Generate 3D Shapes with Sharp Features and Parametric Control?}

\author{Zeqing Yuan\thanks{equal contribution. Zeqing is the project lead. Haoxuan participated in dataset construction and testing.}\\
Zhejiang University 
\And
Haoxuan Lan\footnotemark[1]\\
Zhejiang University 
\And
Qiang Zou\\
Zhejiang University 
\And
Junbo Zhao \\
Zhejiang University 
}

\newcommand{\numData}{57}
\newcommand{\dataset}{3D-PreMise}

\iclrfinalcopy 
\begin{document}

\maketitle

\begin{abstract}
Recent advancements in implicit 3D representations and generative models have markedly propelled the field of 3D object generation forward. However, it remains a significant challenge to accurately model geometries with defined sharp features under parametric controls, which is crucial in fields like industrial design and manufacturing. To bridge this gap, we introduce a framework that employs Large Language Models (LLMs) to generate text-driven 3D shapes, manipulating 3D software via program synthesis. We present \dataset, a dataset specifically tailored for \textbf{3D} \textbf{P}a\textbf{r}am\textbf{e}tric \textbf{M}odeling of \textbf{i}ndustrial \textbf{s}hap\textbf{e}s, designed to explore state-of-the-art LLMs within our proposed pipeline. Our work reveals effective generation strategies and delves into the self-correction capabilities of LLMs using a visual interface. Our work highlights both the potential and limitations of LLMs in 3D parametric modeling for industrial applications.

\end{abstract}

\section{Introduction}
\label{intro}
The integration of software in computational design and manufacturing has revolutionized various engineering sectors, offering significant advancements and efficiencies. However, this integration necessitates a deep understanding of domain-specific knowledge, underscoring the necessity for the development of more accessible interfaces, such as natural language processing, in industrial applications.

In recent years, generative models have made remarkable strides in non-engineering domains. The field of 3D object generation is undergoing a transformative shift, propelled by groundbreaking developments in implicit 3D representation \citep{mildenhall2020nerf, kerbl20233d} and advanced generative models \citep{ho2020denoising}.
While current methods employing implicit representations can generate intricate objects from text and image inputs, they inherently cannot preserve sharp features that contains engineering semantics (e.g. sharp edges, precise dihedral angles) when converting to explicit representations, a critical limitation for applications such as industrial design and manufacturing. For instance, generating a chair with precise dimensions or a plastic bottle cap with parallel and equidistant edges remains a challenge.

This work explores a novel approach to 3D object generation merging the capabilities of LLMs with the precision of 3D software. We harness LLMs to interpret text inputs and generate code to control 3D modeling software like Blender, producing compact 3D shapes with sharp features that adhere to strict parametric controls. This pipeline is advantageous for two reasons:

Firstly, 3D modeling code emerges as an ideal medium for depicting everyday objects, particularly in industrial settings, due to its compactness and high controllability that align well with design intentions. In contrast, implicit 3D representations such as neural radiance fields and Gaussian Splatting, despite their expressiveness for intricate shapes, struggle with maintaining sharp features due to inevitable conversion. Furthermore, directly generating explicit representations, like meshes, currently cannot offer precise parametric controls.

Secondly, advanced LLMs have shown remarkable understanding of 3D spaces and aptitude in programming. They can perform 3D spatial reasoning, planning \citep{sun20233dgpt} and demonstrate high accuracy in programming tasks \citep{zhou2023language}. Therefore, it is promising that, given appropriate prompts and fine-tuning, LLMs can effectively generate 3D objects using modeling code as an intermediary.

Despite its potential, this approach encounters significant challenges. Firstly, the model must adeptly reason about spatial relationships and cohesively assemble object components within a unified coordinate system. Additionally, it entails intricate geometric calculations, including the application of complex trigonometric functions. Lastly, the LLM must possess commonsense reasoning to deduce stable and realistic structures from brief natural language descriptions.

To validate this approach, we introduce \dataset, a dataset comprising test programs and data samples with problem descriptions and corresponding ground-truth code. This dataset focuses on typical industrial objects with exact geometries. We conduct experiments with state-of-the-art LLMs to evaluate their performance and analyze generation strategies.



Our contributions are three-fold:
\begin{itemize}
    \item We introduce a self-refining framework for 3D shape generation with parametric control, which leverages LLMs to control 3D software through code;
    \item We construct a benchmark dataset and conduct experiments to analyze the capacities of cutting-edge LLMs;
    \item We explore effective generation strategies and self-correcting capacity through a multimodal interface.
\end{itemize}

\section{Related Work}
\subsection{Text-driven 3D Object Generation}
The field of Text-driven 3D object generation has witnessed remarkable progress. Clip-Mesh \citep{khalid2022clipmesh} utilizes CLIP guidance for zero-shot 3D generation. DreamFusion \citep{poole2022dreamfusion} proposes a stable paradigm using Score Distillation Sampling (SDS) loss to distill pretrained text-to-image diffusion models and achieves significant improvement. Magic3D \citep{lin2022magic3d} enhances visual quality in high resolution by introducing a two-stage optimization framework. Fantasia3D \citep{chen2023fantasia3d} introduces richer representations. ProlificDreamer \citep{wang2023prolificdreamer} proposes variational score distillation (VSD) to address the over-saturation problem. DreamBooth3D \citep{raj2023dreambooth3d} proposes a 3-stage optimization strategy to jointly leverage 3D consistency of NeRF and the personalization capability of text-to-image models. MVDream \citep{shi2023mvdream} improves the geometric consistency by applying multi-view diffusion model as a 3D prior with Score Distillation. To accelerate generation process, Instant3D \citep{li2023instant3d} proposes a two-stage paradigm that first generates a sparse set of views and then regresses NeRF with a transformer-based reconstructor; DreamGaussian \citep{tang2023dreamgaussian} designs a generative 3D Gaussian Splatting model for faster generation speed.

However, these methods rely on implicit 3D representations, which inherently struggles to retain sharp features embodying engineering semantics during the conversion to practical explicit forms. Consequently, despite their increasing sophistication, these methods fall short in meeting the specific demands of industrial applications. 

MeshGPT \citep{siddiqui2023meshgpt} specializes in creating compact geometries with sharp edges through mesh generation. However, the proposed autoregressive pipeline lacks the ability to support parametric control. 

3D-GPT \citep{sun20233dgpt} introduces a procedural approach, employing LLMs to generate Python code in Blender for 3D scene creation, notably enhancing semantic precision in object modeling. It effectively interprets specific descriptions of a class of objects, producing satisfactory outcomes. However, in the realm of industrial applications, which our research targets, descriptions typically include specific, precise parameters of each object. Moreover, our evaluation emphasizes exact dimensional parameters, in contrast to previous studies that primarily rely on semantic metrics like the CLIP score. Consequently, our work delves into achieving a higher level of precision through parametric control.

\citet{makatura2023large} evaluates the capabilities of GPT-4 in computational design and manufacturing, presenting cases and analysis in shape modeling of LLM using Constructive Solid Geometry. However, the investigation revolves around design abilities given high-level inputs, whereas our work extends to parametric modeling given detailed inputs.

\subsection{Large Language Models for Program Synthesis}
Recent advancements have seen large language models (LLMs) like GPT-4 \citep{openai2023gpt4} and Code Llama \citep{rozière2023code} excel in generating code snippets from natural language inputs. These models have demonstrated remarkable proficiency in writing Python functions on benchmarks such as HumanEval \citep{chen2021humaneval} and MBPP \citep{austin2021mbpp}. In response to their rapid development, more comprehensive benchmarks like ClassEval \cite{du2023classeval}, which assesses class-level Python code, and HumanEval-X \citep{zheng2023codegeex}, which extends to multilingual programming in languages like C++, Java, and JavaScript, have been introduced. Despite these advancements, the specific application of LLMs in 3D object modeling remains under-explored. \dataset\ is the pioneering benchmark dataset focused on program synthesis for 3D object modeling, delving into the nuances of 3D geometric reasoning within programming.  While existing code generation benchmarks evaluate programming logic, \dataset \ investigates further on 3D modeling, which involves spatial reasoning and geometric calculation.


Therefore, the potential of LLMs' capacity in 3D parametric modeling remains largely untapped. To bridge this gap, we develop \dataset\ to scrutinize LLMs and generation strategies, with the goal of catalyzing progression in this area.

\section{Pipeline}
\begin{figure}[H]
    \centering
    \includegraphics[width=1.0\linewidth]{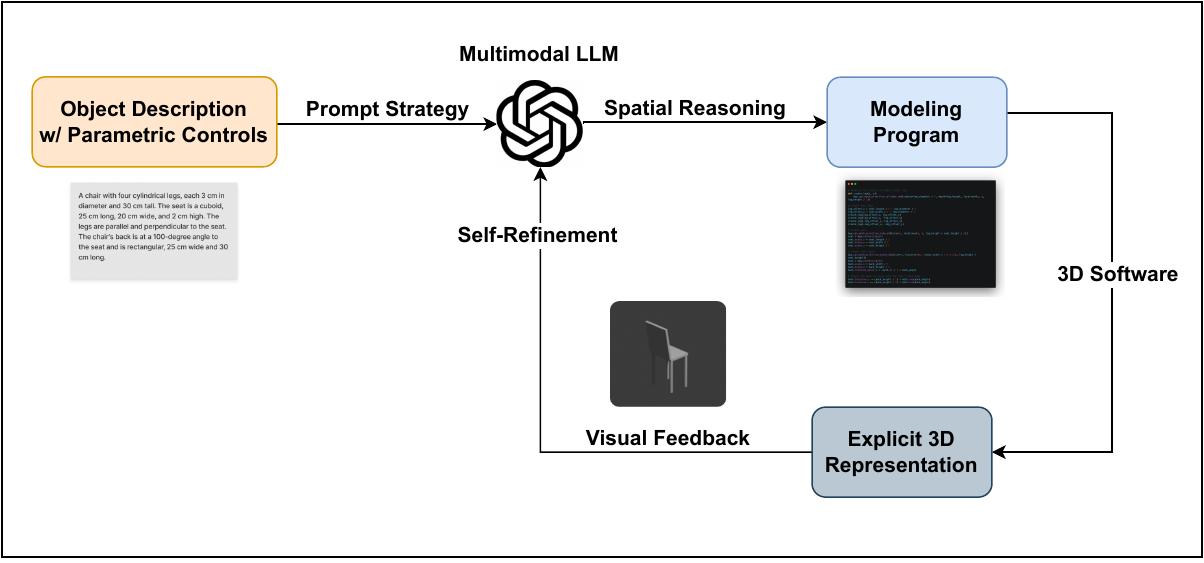}
    \caption{Pipeline Overview}
    \label{fig:pipeline}
\end{figure}

In this section, we introduce a pipeline that integrates LLMs, 3D modeling platforms, and a multimodal interface to facilitate iterative refinement. 
The methodology unfolds across the following stages, as depicted in Figure \ref{fig:pipeline}.

The pipeline takes in text input that precisely defines an object and incorporates strict parametric controls. To augment the LLM's processing and reasoning capabilities, we employ prompting strategies such as in-context learning and chain-of-thought prompting. The LLM then undertakes spatial reasoning to synthesize the modeling programs. These programs are executed on a 3D modeling platform to yield an explicit 3D representation of the object. Subsequently, we render the object to produce visual feedback. This visual output is then fed back into the multimodal LLM, enabling iterative self-refinement to enhance object modeling.



\section{Benchmark Dataset \dataset}
In this section, we elaborate on our proposed benchmark dataset, \dataset.

\subsection{Overview}
Our dataset comprises \numData \ samples, each consisting of two parts: a prompt and a canonical modeling program. The dataset's emphasis is on standalone, rigid objects typical in industrial design settings.

In terms of evaluation, we employ a specialized testing program. 
It analyzes the generated modeling programs and determines the outcome as either 'pass' or 'fail'. This is further quantified by a similarity distance metric that gauges the accuracy of the generated object against the prompt's specifications.

\subsection{Data Samples}
A data sample in \dataset \ contains specific descriptions of an object and a ground-truth modeling program. As an illustration, a sample is shown as follows.
\begin{figure}[H]
    \centering
    \includegraphics[width=0.8\linewidth]{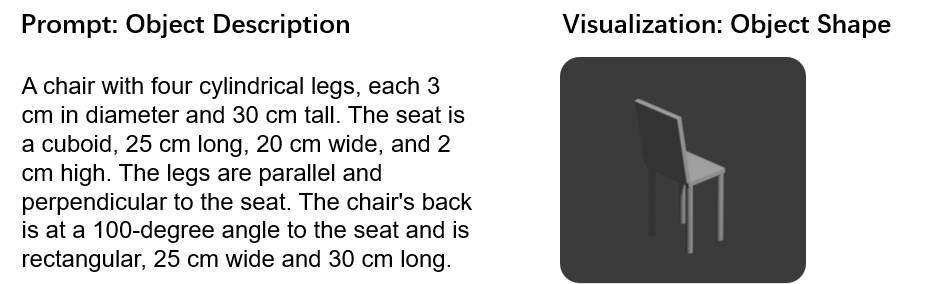}
    \caption{Example object description in \dataset}
    \label{fig:data_sample}
\end{figure}

The prompts in the dataset are meticulously crafted to describe the shape of an object with specific definition detailing parameters.


Correspondingly, the canonical solution presents a correctly formulated Python program using the 'bpy' package for Blender that models the object shape as described in the prompt. 
Each script typically starts with importing necessary libraries, setting up the environment for 3D operations. 
Variables are declared to hold specific measurements, reflecting the dimensions of the object being modeled, such as lengths, widths, heights, and angles. These variables allow for parametric control over the object's geometry and are often accompanied by comments that indicate the real-world size they represent.
Functions are a key component of the code, designed to encapsulate the creation of different parts of the object. For instance, a function might be dedicated to creating a single element like a leg of a chair or a wheel of a vehicle, with parameters that allow for position and size customization.
The main body of the script calls these functions and sets properties to assemble the complete object. It uses transformation operations such as scaling, rotation, and translation to position elements correctly in the 3D space.
Lastly, the code is tailored to interact with specific 3D software APIs through Blender's 'bpy' package. This interaction includes the use of functions to add primitives and modify object properties.

\subsection{Characteristics}

\textbf{Scale and Diversity}. Our dataset consists of \numData \ samples and encompasses a range of categories, such as furniture, toys, and decorative items.
\begin{figure}[h]
    \centering
    \includegraphics[width=0.95\linewidth]{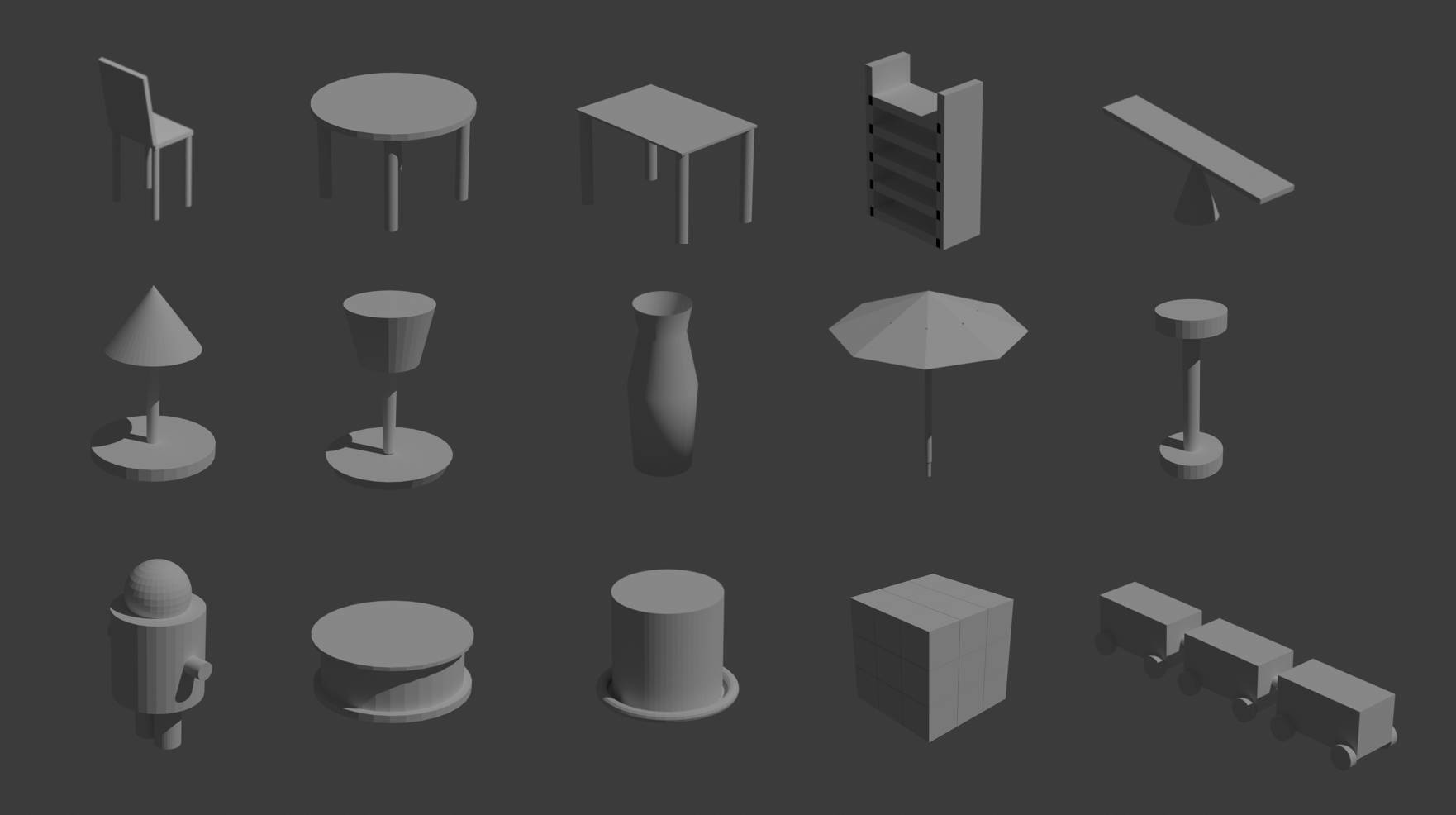}
    \caption{Objects in \dataset\ Dataset}
    \label{fig:enter-label}
\end{figure}

\textbf{Object Geometrical Characteristics.} Our dataset targets objects prevalent in industrial design and manufacturing, characterized by their strictly defined shapes, usually composed of fundamental geometries such as cuboids and cylinders. 

\textbf{Object Description Characteristics.} The prompt descriptions of objects in our dataset are almost fully defined, leaving little space of ambiguity, which emulates the scenarios in industry application. In this way, we aim to enable replacing complex engineering drawings with natural language interface for controllable industrial design.

\subsection{Test Program}
While most works on 3D object generation adopt metrics like Fréchet inception distance (FID), CLIP R-Precision or user studies, our setting targets precise parametric modeling. Therefore, we implement a test program to evaluate if the result is accurately faithful to prompt description.

\subsubsection{Generation Correctness}
The test program first executes the synthesized and ground-truth programs within a virtual sandbox to retrieve the mesh data, $M_S$ and $M_T$, generated using Blender 'bpy' package.
To enhance efficiency, we extract only the vertices to form two point clouds, $P_S$ and $P_T$, for testing, which empirically satisfies our demand.

To eliminate the influence of position and orientation on geometric judgement, we normalize the point clouds into $P_S'$ and $P_T'$ as follows. 

\begin{enumerate}
\item Translation to Origin: 
    Calculate the geometric center \( \mathbf{c} \) of each point cloud and translate all points such that the geometric center moves to the origin. The transformation for a point \( \mathbf{p} \) in point cloud \( P \) is given by:
    \[ \mathbf{p}' = \mathbf{p} - \mathbf{c} \]
    where \( \mathbf{c} = \frac{1}{|P|} \sum_{\mathbf{p} \in P} \mathbf{p} \).

\item Calculation of Principal Axes: 
    Compute the covariance matrix \( \mathbf{C} \) of the translated point cloud. The principal axes are the eigenvectors \( \mathbf{e}_1, \mathbf{e}_2, \mathbf{e}_3 \) of \( \mathbf{C} \), corresponding to its eigenvalues. The covariance matrix is defined as:
    \[ \mathbf{C} = \frac{1}{|P|} \sum_{\mathbf{p}' \in P} (\mathbf{p}' - \bar{\mathbf{p}}')(\mathbf{p}' - \bar{\mathbf{p}}')^T \]
    where \( \bar{\mathbf{p}}' \) is the mean of the translated points.

\item Rotation Transformation: 
    Apply a rotation matrix \( \mathbf{R} \) to align the principal axes with the coordinate axes. The rotation matrix is formed by the eigenvectors as its columns:
    \[ \mathbf{R} = [\mathbf{e}_1 \, \mathbf{e}_2 \, \mathbf{e}_3] \]
    Each point \( \mathbf{p}' \) in the translated point cloud is then transformed as:
    \[ \mathbf{p}'' = \mathbf{R}^T \mathbf{p}' \]
\end{enumerate}

After these transformations, the normalized point clouds \( P_T' \) and \( P_S' \) are obtained, which are independent of their original position and orientation.

Following the normalization, the next step involves comparing the synthesized point cloud $P_S$ with the ground-truth point cloud $P_T$ derived from the canonical solution. This process is executed as follows:

\begin{enumerate}
    \item Point Matching: 
    For each point \( \mathbf{p}_i' \) in the synthesized point cloud (e.g., \( P_S' \)), identify the nearest point \( \mathbf{q}_i' \) in the ground-truth point cloud (e.g., \( P_T' \)). The distance between these two points is calculated using the Euclidean distance formula:
    \[ d(\mathbf{p}_i', \mathbf{q}_i') = \sqrt{(\mathbf{p}_i' - \mathbf{q}_i') \cdot (\mathbf{p}_i' - \mathbf{q}_i')} \]
    
    \item Success Criteria: 
    The match is considered successful if the distance \( d(\mathbf{p}_i', \mathbf{q}_i') \) does not exceed a predefined threshold \( \delta \):
    \[ \text{Match Success} \Leftrightarrow d(\mathbf{p}_i', \mathbf{q}_i') \leq \delta \]

    \item Reverse Matching: 
    To ensure the robustness of the match, a reverse matching process is also conducted. In this process, each point in the ground-truth point cloud is matched to the nearest point in the synthesized point cloud, following the same distance criteria.

    \item Final Decision:
    If all the points in both point clouds successfully match with their counterparts under these criteria, the test program outputs a 'pass'. Otherwise, it outputs a 'fail'.
\end{enumerate}

\subsubsection{Deviation Distance}
To further quantify the deviation between the generated code and ground-truth solution, we introduce Chamfer distance on vertex location, defined as follows, where $V_1$ and $V_2$ are two point clouds to be compared.

$$
d_{CD}(P_S', P_T') = \cfrac{1}{|P_S'|}\sum\limits_{x \in P_S'}\min\limits_{y \in P_T'}||x-y||_2^2 + \cfrac{1}{|P_T'|}\sum\limits_{x \in P_T'}\min\limits_{y \in P_S'}||x-y||_2^2
$$

\section{Experiments}
Using \dataset, we conduct experiments and analysis with state-of-the-art LLMs on 3D shape generation through modeling program synthesis. 

\subsection{Generation Strategies and Prompt Design}
In this section, we explore the impact of various generation strategies on the performance of LLMs within the scope of \dataset. Our experiment is structured around the following strategies:

\begin{itemize}
    \item Zero-Shot Baseline: Directly input the task requirement and description.
    \item Zero-Shot Chain-of-Thought: Instruct the LLM to answer progressive questions and think step by step before generating code. The questions include listing specific shape, parameters and spatial position of each element. 
    \item One-Shot In-Context Learning: Provide an example containing a description and an answer.
    \item Few-Shot In-Context Learning: Provide 3 examples containing descriptions and answers.
    \item One-Shot Chain-of-Thought: Provide an example containing description, reasoning and answer. In addition, instruct the LLM to answer progressive questions before generating code.
\end{itemize}

\subsection{Implementation Details}
We adopt the OpenAI API for model \citep{openai2023gpt4} interface in December, 2023. 
The maximum window size is set to 1024 tokens.
As for inference settings, we follow the mainstream strategies in recent code generation works \citep{du2023classeval} : 
\begin{enumerate}
    \item Greedy Sampling: setting the temperature as 0 to generate one greedy sample and calculate $Pass@1$ metric.
    \item Nucleus Sampling: setting the temperature as 0.9 to generate 3 samples and calculate $Pass@k$, where $k={1,3,5}$. 
\end{enumerate}

\subsection{Quantitative Results}
The experiment results are quantified using the common metric $pass@k$ \citep{spoc2019}:
$$
pass@k = \mathbb{E}_{\text{Problems}} \left[ 1 - \frac{{n-c \choose k}}{{n \choose k}} \right]
$$

The results under different settings are shown below.
\begin{table}[h!]
\centering
\caption{Experiment Results on \dataset}
\begin{tabular}{l|l|c|ccc}
\hline
\multirow{2}{*}{Model} & \multirow{2}{*}{Generation Strategy} & Greedy Sampling & \multicolumn{3}{c}{Nucleus Sampling} \\ \cline{3-6} 
                &                   & Pass@1      & Pass@1       & Pass@3       & Pass@5      \\ \hline
GPT-4           & Zero-Shot         & 7.0\%       & 3.5\%        & 5.3\%        & 7.0\%      \\
GPT-4           & Zero-Shot CoT     & 5.3\%       & 5.3\%        & 7.0\%        & 7.0\%      \\
GPT-4           & One-Shot          & 12.3\%      & 12.3\%       & 14.0\%       & 14.0\%      \\
GPT-4           & Few-Shot          & 12.3\%      & 12.3\%       & 14.0\%       & 15.7\%      \\
GPT-4           & One-Shot CoT      & 17.5\%      & 15.8\%       & 19.3\%       & 19.3\%      \\\hline
\end{tabular}
\label{tab:my_label}
\end{table}

According to the quantitative results, we can summarize the following findings.
\begin{itemize}
    \item \textbf{Finding 1}: In-context learning enhances the LLM performance. However, the performance gap between one-shot and few-shot scenarios is marginal. This observation aligns with expectation, considering the difficulty in deciphering the internal process of analysis between description and modeling code.
    \item \textbf{Finding 2}: The adoption of chain-of-thought prompting notably advances the performance. Nonetheless, the inclusion of an explicit example of the analysis process is essential for this improvement. The provision of guiding questions alone does not yield any measurable enhancement.

\end{itemize}

\subsection{Case Analysis}
In this section, we examine the incorrect cases found in our experiments. To better understand these issues, we've grouped the errors into several categories.

\begin{itemize}
    \item \textbf{Syntax Error}: issues related to the modeling code, which may prevent correct execution.
    \item \textbf{Geometric Error}:
    \begin{itemize}
        \item \textbf{Structural Configuration Error}: the arrangement of geometric elements fundamentally contradicts the description.
        \item \textbf{Spatial Precision Error}: minor inaccuracies in spatial parameters, including slight deviations in the position, scale, or rotation of specific elements.
    \end{itemize}
    \item \textbf{Logical Error}: errors that arise from a lack of adherence to real-world logic or commonsense principles.
\end{itemize}

\begin{figure}[H]
    \centering
    \includegraphics[width=1.0\linewidth]{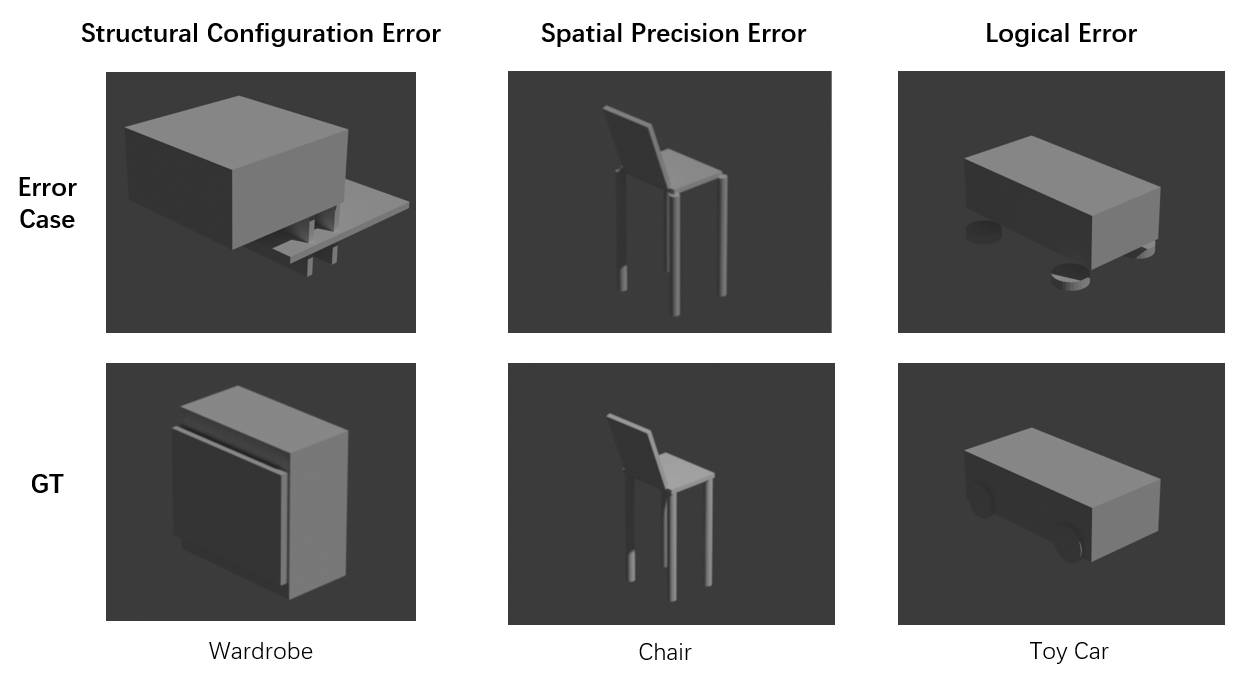}
    \caption{Illustration of Error Categories}
    \label{fig:enter-label}
\end{figure}

To navigate the problem, we conduct a detailed statistical analysis of the frequency in which each error category occurred. The experiment is based on GPT-4 model with temperature set to $0.9$.

\begin{figure}[H]
    \centering
    \includegraphics[width=1.0\linewidth]{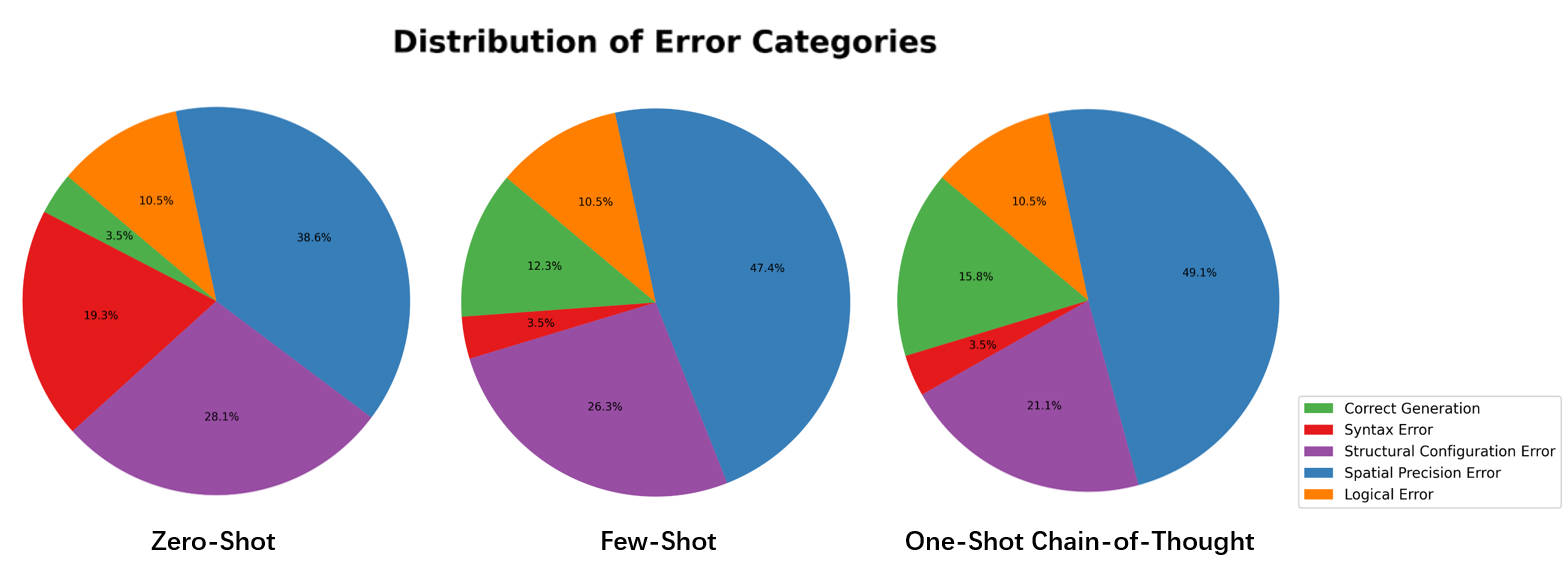}
    \caption{Statistical Analysis on Failure Cases}
    \label{fig:enter-label}
\end{figure}

The analysis of error distributions leads to the following findings.

\begin{itemize}
    \item \textbf{Finding 1:} The generation suffers from spatial precision error most frequently, which is aligned with our expectation that precise control is the painpoint for 3D shape generation.
    \item \textbf{Finding 2:} For cutting-edge LLMs like GPT-4, appropriate in-context learning strategy can significantly alleviate syntax error, making modeling program a feasible representation.
    \item \textbf{Finding 3:} Chain-of-Thought strategy proves to be effective in improving LLM's structural arrangement when modeling objects.
\end{itemize}

\begin{figure}[H]
    \centering
    \includegraphics[width=1.0\linewidth]{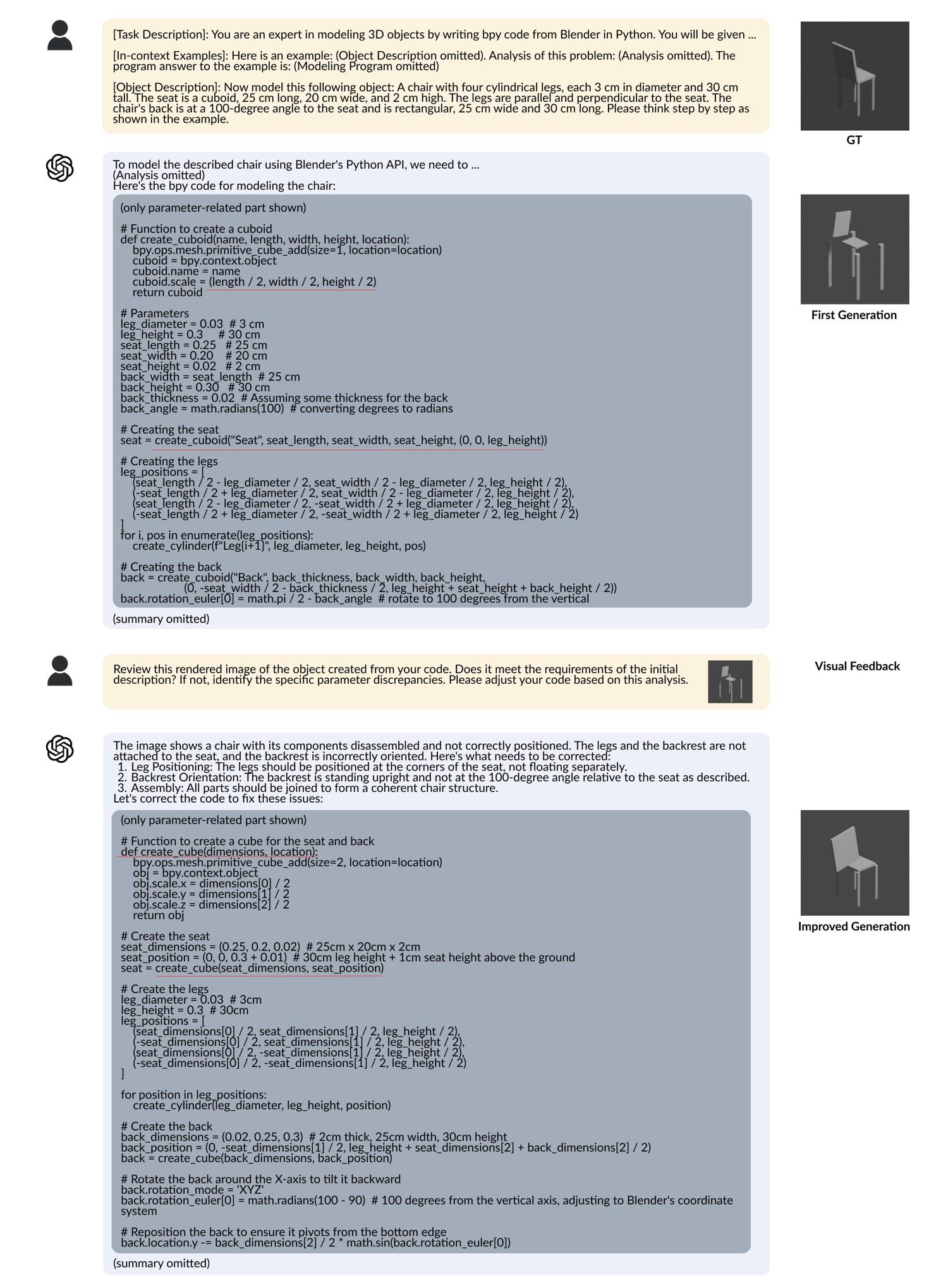}
    \caption{\textbf{Self-Improving with Visual Feedback}. Demonstration of GPT-4's ability to self-correct modeling programs via its visual interface. The red line highlights the modification that resolved the issue with the seat's scale.}
    \label{fig:self-improve}
\end{figure}

Further analysis at the program level of failure cases in parametric control has revealed the following key insights:
\begin{itemize}
    \item \textbf{Finding 1}: The most frequent errors stem from incorrect programming practices, notably the redundant scaling applied both within function definitions and function parameters.
    \item \textbf{Finding 2}: Parametric error rates diminish when parameters described in the prompt are explicitly defined as variables in the code. This practice not only enhances the code's readability but also facilitates more effective reasoning.
\end{itemize}

\subsection{Evaluating Self-Enhancement in Multimodal LLM via Visual Feedback}
In this section, we examine how the cutting-edge multimodal LLM, GPT-4, can self-improve its modeling program from visual feedback.
We use the OpenAI ChatGPT interface to interact with the GPT-4 model in December, 2023.

We employ a one-shot chain-of-thought strategy to input the requirements and description. The LLM-generated code is then rendered in Blender, with the resulting image fed back into GPT-4's visual interface for self-correction of the modeling program. This iterative process is used to scrutinize and refine the generated code.

As illustrated in Figure \ref{fig:self-improve}, GPT-4 initially erred in creating the cuboids for the seat and back. It developed a function that inadvertently halved the scale based on the input length and width, leading to a disproportionately small seat that did not align with the chair's legs. Upon reviewing the visual feedback from the rendered result, GPT-4 modified the function by adjusting the size parameter in the 'primitive\_cube\_add' function from 1 to 2. While this approach to coding was unconventional, it effectively corrected the initial mistake, demonstrating that the LLM was capable of identifying and addressing the error.

Our expanded experiments yielded several insights regarding the self-correcting abilities of LLMs through multimodal interfaces:
\begin{itemize}
    \item \textbf{Finding 1}: GPT-4 is able to rectify obvious parametric mistakes arising from incorrect programming practices.
    \item \textbf{Finding 2}: GPT-4 shows a lack of sensitivity to common sense errors discernible from visual feedback, such as a cylindrical chair leg incorrectly positioned partly outside, rather than fully beneath, the seat.
    \item \textbf{Finding 3}: The self-correcting capabilities of GPT-4 are notably limited in scenarios involving complex mathematical calculations in geometry. (e.g. they struggle to correct errors stemming from the misapplication of trigonometric functions like sine and cosine.)
    \item \textbf{Finding 4}: GPT-4 hallucinates in response to the visual feedback, sometimes offering contradictory statements.
\end{itemize}

\section{Conclusion and Future Work}
In this study, we have developed a pipeline designed for the generation of 3D shapes with parametric controls and engineering semantics, harnessing the power of multimodal LLMs to utilize 3D software through program synthesis. We introduced \dataset, a comprehensive dataset supported by a specialized testing program, to critically assess the capabilities of LLMs within this innovative context. Our investigation into various generative strategies has identified key techniques that significantly enhance model performance in different dimensions. We have also explored the effectiveness of a visual interface in augmenting the self-correction abilities of LLMs. Our experiments and analysis have revealed the capacities of LLMs in spatial reasoning, geometric computing, program synthesis and multimodal self-correction.



In the future, we plan to enlarge our dataset and apply efficient fine-tuning to improve our pipeline. Specifically, we aim to enhance LLM's capabilities in spatial computation, geometric commonsense and the capacity for self-refining 3D modeling programs based on visual feedback, emulating human cognition.

\section*{Acknowledgments}
We would like to thank Kenan Yu and Jiaying Lai for their contribution to dataset construction and insightful discussion.

\bibliography{main}
\bibliographystyle{unsrtnat}


\end{document}

%% file: math_commands.tex
\usepackage{amsmath,amsfonts,bm}

\def\eqref#1{equation~\ref{#1}}

\def\1{\bm{1}}

\DeclareMathAlphabet{\mathsfit}{\encodingdefault}{\sfdefault}{m}{sl}
\SetMathAlphabet{\mathsfit}{bold}{\encodingdefault}{\sfdefault}{bx}{n}

